\documentstyle {article}

\begin{document}

\begin{flushright}
	RU-97-40
\end{flushright}

\vspace{2cm}

\begin{center}
\Large
 {\bf Matrix Model for the Transition Expansion}
 {\bf	of Dispersive Corrections to}
 {\bf 	Nucleon-Nucleus Total Cross Sections}

\vspace{1cm}

	David R. Harrington

\vspace{.4cm}

	{\em Department  of Physics and Astronomy}

	{\em Rutgers University}

	{\em P.O. Box 849, Piscataway, NJ 08855-0849}
\end{center}

\vspace{1cm}
\normalsize

\begin{center}
	Abstract
\end{center}
A finite-dimensional matrix model for the nucleon-nucleon cross section operator is used
to calculate the dispersive correction to nucleon-nucleus total cross
sections, and the leading terms in its expansion in the number of
inelastic transitions in the high-energy limit where the longitudnal
momentum transfers can be ignored.  Results for matrices of different
dimensions, but
giving the same second and third  cross section moments, are compared
for scattering from $ ^{208}Pb$.  The leading second order terms in the
expansion are  accurate to better than 10\% , but
always  larger than the exact result.  The  3rd order terms
overcorrect, but are nearly cancelled by the 4th  order terms.  The 5th
and 6th  order terms are smaller, and also tend to cancel, but not to the
degree of the 3rd and 4th order terms. The series converges more
rapidly for smaller matrix dimensions, and in each order the
magnitudes of the terms increase with the dimension of the matrix.

\newpage

\begin{center} {\bf I Introduction}
\end{center}

	One consequence of the composite structure of nucleons is a
decrease in nucleon-nucleus total cross sections due to transitions
between different internal states of the  projectile nucleon.  This
decrease can be calulated \cite{bla,har} using an operator to represent the
nucleon-nucleon total cross section, with the matrix elements
representing the probability amplitudes for forward scattering transitions between different
states of the nucleon.

	If the longitudnal momentum transfer due to the different
masses of these states cannot be ignored  this dispersive
correction can only be calculated by expanding in terms of the number of
inelastic transitions between states of different mass.  Because of
the difficulty of this calculation, and uncertainties in the nature of
the cross section operator, only the leading term in this expansion
has been evaluated \cite{jen}, and then only approximately. (A closely
related effect in quasi-elastic electron scattering can complicate the
analysis of color transparency effects in these reactions \cite{kop}.)
In this paper we take a first step toward testing the convergence of
this expansion by using finite matrix models for the cross section operator
and taking the high energy limit in which the longitudnal momentum
transfer vanishes.

	Section II reviews the formulas and notation expressing the
dispersive corrections to the hadron-nucleus total crosss sections in
terms of an integral over the generating function for the cross
section probability distribution function.  The transition expansion
is then reviewed in Sect. III, where it is shown that  each term in
the expansion can be represented as 
a sum over products of transition amplitudes weighted by a function of
the differences among cross sections in the different nucleonic
states.  In Sect. IV the cross section operator is represented by a
finite dimensional matrix depending on two parameters, one of which
fixes the second moment of the operator, the other the third moment.
With this simple form for the cross section operator one can find
relatively simple
analytic expressions for the low order terms in the transition
expansion. These are evaluated and the results presented in
Sect. V.  The results are summarized and discussed in the concluding
Sect. VI.

\begin{center} {\bf II Hadron-Nucleus Cross Sections}

\end{center}

	In this paper it is assumed that the energy of the incident
hadron is high enough that the longitudnal momentum transfers
associated with inelastic scattering can be ignored.  Nucleon overlap
in the target nucleus is also ignored, along with nuclear
correlations.  For heavy nuclei the total cross section for the
scattering of an incident hadron from a nucleus of nucleon number $A$ is
then given by

\begin{equation}	
	\sigma (A) = \sigma ^{G}(A) -\sigma ^{D} (A) ,
\end{equation} 
where

\begin{equation}
	\sigma ^{G,D} (A)=2 \int d^{2}b\,  G^{G,D}(t(A,b)),
\end{equation}
 with the Glauber contribution

\begin{equation}
	G^{G}(t)=1-exp(-t)
\end{equation}
and the ``dispersive'' correction due to diffractive excitation

\begin{equation}
	G^{D}= <1|exp(-\hat{x} t)|1>-exp(-t).
\end{equation}
Here the expectation value is in the first, or ground, mass eigenstate
 of the projectile.  The dimensionless thickness
function $t$ at impact parameter $b$ is given by

\begin{equation}
	t(A,b)=A\sigma T(A,b)/2,
\end{equation}
where $\sigma$ is the projectile-nucleon total cross section and
$T(A,b)$ is the usual thickness function, i.e. the integral of the
nuclear density, normalized to unity, along a straight line at
constant impact parameter.  The dimensionless thickness function
decreases from a maximum value of approximately $(3\sigma /4\pi r_{0}^{2})A^{1/3}$  at zero impact
parameter to zero for impact parameters well outside the nuclear radius
$R\approx r_{0}A^{1/3}$, with $r_{0}\approx 1.14fm$.  Below the values
$A=208$ and $\sigma = 39.8 mb$ are used, giving the maximum value of
$t$ as about 4.3.
The  detailed shape of $t$ as a function of $b$ depends on the shape of the nuclear density: below a
Woods-Saxon form will be assumed.

	In Eqn. 4 the operator
\begin{equation}
 \hat{x} =\hat{\sigma}/<1|\hat{\sigma}|1>
\end{equation}
 is the dimensionless cross section
operator.  The matrix elements of the cross section operator $\hat{\sigma}$ itself give the
cross sections for transitions among the various mass eigenstates of
the projectile.  Thus
\begin{equation}
 \sigma=<1|\hat{\sigma}|1>
\end{equation}
 is the total cross
section for the projectile to interact with a single nucleon, while
, assuming all forward amplitudes are pure imaginary, 
the total forward cross section for diffraction dissociation is
\begin{eqnarray}
d\sigma /dt|_{t=0} & = & \pi \sum_{j\not=1}|<j|\hat{\sigma}/4\pi |1>|^{2} \\
			& = & \sigma ^{2} [<1|\hat{x} ^2|1> -
1]/(16\pi ) .	
\end{eqnarray}

	Since there {\em is} diffraction dissociation the operators
$\hat{\sigma}$ and $\hat{x}$ are clearly not diagonal in the space of
mass eigenstates.  They can, however, be diagonalized using their
eigenstates $|\alpha )$ , where

\begin{equation}
	\hat{x} |\alpha ) = x_{\alpha} |\alpha ).
\end{equation}
Then the generating function for the probability distribution, needed
to evaluate Eqn. 4, is

\begin{equation}
	<1|exp(-\hat{x} t)|1> = \sum_{\alpha} P_{\alpha}\,
exp(-x_{\alpha} t),
\end{equation}
where
\begin{equation}
	P_{\alpha}=|<1|\alpha )|^{2}
\end{equation}
is the probability of finding the projectile in the scattering eigenstate
$|\alpha )$ when it is in the mass ground state $|1>$.  Of course $\hat{x}$
will have in general a continuous as well as a discrete spectrum, so
that the sum in Eqn. 11 should be interpreted as a sum plus an
integral.

	In this paper, however, $\hat{x} $ will be represented by a
finite dimensional matrix determined by two parameters, chosen to
conform to the following constraints:
\begin{enumerate}
	\item The off-diagonal elements $<i|\hat{x} |j>$ should
decrease as $|i-j|$ increases .
	\item The diagonal elements $<j|\hat{x} |j>$ should
increase, or at least not decrease, as j increases.
\end{enumerate}
Both of these requirements are suggested by non-relativistic wave
function models: the overlap of the two wavefunctions should decrease
with increasing quntum number separation, and the spatial extent of
the wavefunctions should increase with increasing quantum
numbers. (Related work using finite matrices to describe inelastic
scattering can be found in \cite{fra,huf,won}.)

	Although the longitudnal momentum transfer is ignored in the
high energy limit used here, it would be reasonably easy to include it
along the lines of \cite{jen} if the masses of the excited states were
known.   This is in fact one of the motivations for
 developing a finite matrix model.

\begin{center} {\bf III Transition Expansion }

\end{center}

	The operator  $\hat{x} $ can be separated into two components, one
diagonal and the other off-diagonal (inelastic) in the mass eigenstates:
\begin{equation}
	\hat{x} = \hat{x}_{d} + \hat{x}_{I} ,
\end{equation}
where
\begin{equation}
	<i|\hat{x}_d |j> \equiv \delta _{ij} <i|\hat{x} |i> .
\end{equation}
Then, just as in the case of time-dependent perturbation theory,
\begin{equation}
	\hat{U} (t) \equiv exp(-\hat{x} t)
\end{equation}
can be expanded in powers of $\hat{x} _I $ :
\begin{equation}
	\hat{U} (t) = e^{-\hat{x}_{d} t} [1+ \sum_{n=1}^{\infty}
\hat{U}_{I} ^{(n)}(t) ],
\end{equation}
where the $nth$ term is given by the t-ordered integral
\begin{equation}
	\hat{U}_{I} ^{(n)}(t) = (-1)^{n}\int_{0}^{t} dt_{n}\ldots dt_{1}
\theta (t_{n}-t_{n-1})\ldots \theta (t_{2}-t_{1})
\hat{x} _{I}(t_{n})\ldots  \hat{x} _{I}(t_{1}) , 
\end{equation}
with
\begin{equation}
	\hat{x} _{I}(t) \equiv e^{ \hat{x}_{d}t} \hat{x} _I e^{-\hat{x} _{d}t} .
\end{equation}
The diffractive contribution to $G(t)$ can then be expanded in powers
of $\hat{x}_{I}$, which is equivalent to an expansion in the number of
transitions between different mass eigenstates:
\begin{equation}
	G^{D}(t) = \sum_{n=2}^{\infty} G^{D(n)}(t),
\end{equation}
where
\begin{equation}
	G^{D(n)}(t) = e^{-t} <1| \hat{U}_{I}^{(n)}(t)|1>.
\end{equation}
Inserting complete sets of mass eigenstates, this can be written as
\begin{eqnarray}
	G^{D(n)}(t) & = & [(-t)^{n} e^{-t}/n!]\sum_{j_{1},\ldots ,j_{n-1}}
<1|\hat{x}_{I} |j_{n-1}>\ldots <j_{1} |\hat{x}_{I} |1> \nonumber \\
                    &   & f^{(n)}(y_{j_{n-1}},\ldots ,y_{j_{1} }),
\end{eqnarray}
where
\begin{equation}
	y_{j} = (x_{jj}-1)t
\end{equation}
and the functions $f^{(n)}$ are defined as the  ordered integrals
\begin{eqnarray}
	f^{(n)}(y_{n-1}, \ldots ,y_{1} ) & \equiv &  n! \int_{0}^{1} du_{1}
 \ldots du_{n}
 \theta (u_{n} - u_{n-1}) \ldots \theta (u_{2} - u_{1} ) \nonumber \\ 
                                  &    & e^{-y_{n-1} (u_{n} - u_{n-1})} \ldots
 e^{-y_{1} (u_{2} -u_{1}) }.
\end{eqnarray}
These symmetric functions are normalized to equal one when all arguments vanish,
and are monotonically decreasing functions of each argument.

The leading term in this expansion for $G^{D}$ is well known and often used to
estimate $G^{D}$, especially when the longitudnal monentum transfers
are not negligible \cite{jen}:
\begin{equation}
	G^{D(2)}(t) = (t^{2}/2)e^{-t} \sum_{j\neq 1}
<1|\hat{x}|j><j|\hat{x} |1> f^{(2)} ((x_{jj}- 1)t),
\end{equation}
where
\begin{equation}
	f^{(2)}(y) = 2(e^{-y}-1+y)/y^{2}.
\end{equation}
One in principle needs to know all matrix elements of
$\hat{x}$ to evaluate $G^{D(2)}$ and higher order terms, although in practice the sum over
$j$ may converge rapidly.
 
	If all diagonal matrix elements of $\hat{x}$ are equal then
$\hat{x}_{d} = 1$, all $y_{j} 's$ vanish, and the above expressions simplify to
\begin{equation}
	G^{D(n)} = <1|\hat{x}_{1}^{n}|1> (-t)^{n} e^{-t}/n! ,
\end{equation}
where $\hat{x}_{1} \equiv \hat{x} - 1$ is the shifted operator.

\begin{center} {\bf IV Matrix Model }

\end{center}

	The dimensionless cross section operator $\hat{x}$ determines
$G^{D}(t)$ and its expansion in inelastic transitions (assuming that
the mass eigenstates are known).  As mentioned above, $\hat{x}$ may in
general have a
discrete and/or a continous spectrum.  (The parameterizations of
\cite{bla}, for example, are based upon a purely continuous
spectrum for $\hat{x}$.)  Here it is assumed, however, that this operator can
be approximately represented by a  two-parameter matrix of finite
dimension $N$ in the space of
(discrete) mass eigenstates:
\begin{equation}
<i|\hat{x}|j> = \delta _{ij}(j-1)d + f^{|i-j|},
\end{equation}
so that the diagonal elements $ 1, 1+d, 1+2d, \ldots  $ increase
linearly with $i=j$ if $d>0$ while the off-diagonal elements decrease geometrically with
$|i-j|$ if $0<f<1$.  Since
\begin{equation}
	<1|\hat{x}^2 |1>=1 + f^2 + f^4 + \ldots + f^{2(N-1)} =(1-f^{2N})/(1-f^2),
\end{equation}
 the parameter $f$ is fixed
by $<1|\hat{x}^2 |1> $, which is in turned fixed by forward diffaction
dissociation according to Eqn. 9.  The second parameter $d$ can then be
determined from
\begin{equation}
	<1|\hat{x}^3 |1> -<1|\hat{x}^2 |1>= (2+d)[f^2 + 2f^4+ \ldots +(n-1)f^{2(N-1)}] .
\end{equation}
For any $N$,  $f$ and $d$ can therefore be chosen to match the second and
third moments  of $\hat{x} $ \, (the only constraint being that $d$
should  be non-negative):  all higher moments and $ G^D (t)$ are then
completely determined.
	
This parameterization of $\hat{x}$ is rather arbitrary and leads to
what should be considered as a ``toy''  model.  Its main
advantages are that it has enough flexibility to match any desired
$<x^2 >$ and $<x^3 >$ (with the constraint noted above) and that the
lower order terms in the transition expansion are relatively easy to
evaluate.  One can think of many other models with these properties:
to choose among the possibilities one would require a rather
complete realistic model for the composite structure of the nucleon.

	In this model the product of matrix elements in each term in
the sum in Eqn. (21) is just the  parameter $f$ raised to some integer
power.  This multiplies a function $f^{(n)}$ which is a simple symmetric
analytic function of its $n-1$ arguments.  Starting with $f^{(2)}$,
given by Eqn. (25), these functions can be calculated from recursion
relations and power series:
\begin{eqnarray}
	f^{(n+1)}(y_{1} ,y_{2} ,\ldots ,y_{n} ) & = &
-(n+1)[f^{(n)}(y_{2},y_{3},\ldots ,y_{n})-   \nonumber \\
                     &   &  f^{(n)}(y_{1},y_{3},\ldots
,y_{n})]/(y_{2}-y_{1}) \nonumber  \\
          & = & 1-(y_{1}+y_{2}+\ldots +y_{n})/(n+2) + \ldots \, .
\end{eqnarray}

\vspace{.4cm}

\begin{center} {\bf V Evaluation for Nucleon-$^{208}Pb$ Total Cross Section }

\end{center}

	The formulas above can be used to calculate the dispersive
corrections to any total cross section and their expansions in terms of
the number of inelastic transitions for cross section matrices of any
dimension.    Here only the example of nucleons on the  heavy
nucleus $^{208}Pb$ will be considered, taking 39.8 mb for the
nucleon-nucleon total cross section $\sigma$ and using a Woods-Saxon
density with radius $R=6.75 fm$ and surface thickness parameter
$a_{0}=2.3 fm$ for the $^{208}Pb$ nucleus .  (These parameters give an un-corrected Glauber cross
section of 3022 mb.) Here the values $<x^{2}> = 1.25$, as in Ref. 1, and $<x^{3}> = 1.9$, which is close to the
smallest value giving non-negative values of the parameter $d$
according to Eqn. 29, are used for the 2nd and 3rd moments. (Ref. 1 takes  $<x^{3}> = 1.75$ based upon an
approximate analysis of nucleon-deuteron inclusive diffractive scattering.)

 The exact  integrands  $4 \pi b
G^{D}$  in these models, along with the contributions from terms of
order $n=2,3,\ldots ,6$ in inelastic scattering, are shown in Figs. 1
and 2 for matrix  dimensions $N=3$ and $N=6$,
repectively.   The contributions from different $n's$ alternate in sign,
as required by Eqn. 21, with the positive even terms  comparable in
magnitude to the preceding negative odd terms.  This can be understood by
considering the lowest order terms in powers of the
parameter $f^{2}$, which  is never much larger than 0.2.
  A simple analysis shows that  the leading terms in the
$G^{D(3)}$ and $G^{D(4)}$ are both of order $f^{4}$, while the leading
terms in $G^{D(5)}$ and $G^{D(6)}$ are both of order $f^{6}$.
Furthermore, the larger the dimension of the matrix the more terms are
included in the sum in Eqn. 21, and thus the larger the magnitude of
$G^{D(n)}$. The additional terms are in general of higher order in
$f^{2}$, however, so this increase converges for  large N.

 The results for the cross
section corrections after integration over impact parameter are shown
in Table 1.  These results reflect the behavior of the integrands: The
$n=2$ term is accurate to better than   10\%, while the 3rd and 4th order terms range from about 20 to 40\% of
 the exact result, but cancel to better than 4\%. The 5th and 6th order terms are
smaller, and also of opposite sign, but do not cancel to the same
degree. The sum of the $n=2$ through $n=6$ terms in the expansion
gives results which are accurate to  better than 1\% for N=3, but only
to about 5\% for N=6. (The continuous distribution models of Ref. 1  give somewhat
higher values, ranging from 199 to 232 mb, for the dispersive correction to
the total nucleon-$^{208}Pb$ cross sections.)

	The calculations above have been repeated for larger values of
$<x^{3}>$ up to 2.5.  The exact value for the dispersive correction
and the magnitudes of  terms in the
expansion all increase with the dimension of the matrix and decrease
with increasing $<x^{3}>$ (presumably because the diagonal elements of
the matrix become relatively larger as $<x^{3}>$ and therefore
$d$ increase).  The near-cancellation of the 3rd and 4th order terms
seems to be general, so that including these terms never improves the
accuracy significantly. 

\begin{center} {\bf VI Conclusion }

\end{center}

	For the particular matrix models used above, the lowest order
term in the transition expansion for the diffractive correction to the
total cross section gives fairly good  accuracy, but is always a bit
too large.  Including higher
order terms does not improve the accuracy significantly, and in fact
adding only the next-to-leading 3rd order term always gives
poorer accuracy, with a correction less than the exact result.

	It would be interesting to repeat these calculations for other
matrix models to see how general these results are, and to include the
longitudnal momentum transfers which appear for finite energies.
(Preliminary calculations indicate that for constant nuclear densities
the formulas above can still be  used, except that the arguments
$y_{j}$ in Eqn. 21 become complex, with imaginary parts depending on
the nucleon's mass spectrum.)
Including the longitudnal momenta, for example, might change the
phases of the terms in the expansion enough to interfere with the
cancellation of the 3rd and 4th order terms.

\newpage

\begin{center}
{ \bf FIGURES}
\end{center}
\vspace{.4cm}

FIG. 1  The integrands $4\pi bG^{D}$ for the dispersive corrections to the
nucleon-$^{208}Pb$ total cross section using a 3-dimensiional matrix
model as described in the text, with $<x^{2}>=1.25$ and $<x^{3}>=1.9$.
The solid curve is the exact result, the labelled dashed curves the
contributions of order 2 thorugh 6 in inelastic transitions.

\vspace{.2cm}

FIG. 2  The same as FIG. 1, but using a 6-dimensional matrix model.

\vspace{1cm}

\begin{center}
{ \bf TABLE}
\end{center}
\vspace{.4cm}

TABLE 1  The dispersive corrections to the nucleon-$^{208}Pb$ total
cross sections for matrix models of different dimensions, including
the exact results, the contributions from orders 2 through 6 in
inelastic transitions, and the sums  of these five contributions.

\vspace{.5cm}

\begin{tabular} {||c||c|c|c|c|c|c|c||}  \hline
  &  \multicolumn{7}{c||}{Dispersive Corrections to Total Cross
Sections (mb)} \\ \cline{2-8}
Matrix &  &  \multicolumn{6}{c||}{Order in Inelastic Transitions}
\\ \cline{3-8}
 Dimension  &   Exact   &  2 & 3 & 4 & 5 & 6  & Sum  \\ \hline\hline
 3 & 137.64 & 143.41 & -30.37 & 29.46 & -8.51 & 4.58 & 138.56 \\ \hline
 4 & 143.08 & 154.82  & -47.13 & 47.75 & -20.80 & 11.80 & 146.44  \\ \hline
 5 & 145.18 & 158.53 & -53.21 & 57.02 & -29.60 & 18.37 & 151.12 \\ \hline
 6 & 145.82 & 159.58 & -54.95 & 60.29 & -33.61 & 22.13 & 153.43 \\\hline

\end{tabular}


\begin{thebibliography}{99}

\bibitem{bla} B. Bl\"{a}ttel, G. Baym, L. L. Frankfurt, H. Heiselberg,
and M. Strikman, Phys. Rev. D  {\bf 47}, 2761 (1993).

\bibitem{har} D. R. Harrington, Phys. Rev. C {\bf 52}, 926 (1995).

\bibitem{jen} B. K. Jennings, G. A. Miller, Phys. Rev. C {\bf 49},
2637 (1994).

\bibitem{kop} B. Kopeliovich and J. Nemchik, Phys. Lett. B {\bf 368},
187 (1996).

\bibitem{fra} L. Frankfurt, W. R. Greenberg, G. A. Miller, and
M. Strikman, Phys. Rev. C {\bf 46}, 2547 (1992).

\bibitem{huf} J. H\"{u}fner and B. Kopeliovich, Phys. Rev. Lett. { \bf
76}, 192 (1996).

\bibitem{won} C. W. Wong, Phys. Rev. D {\bf 54}, R4199 (1996).



\end{thebibliography}
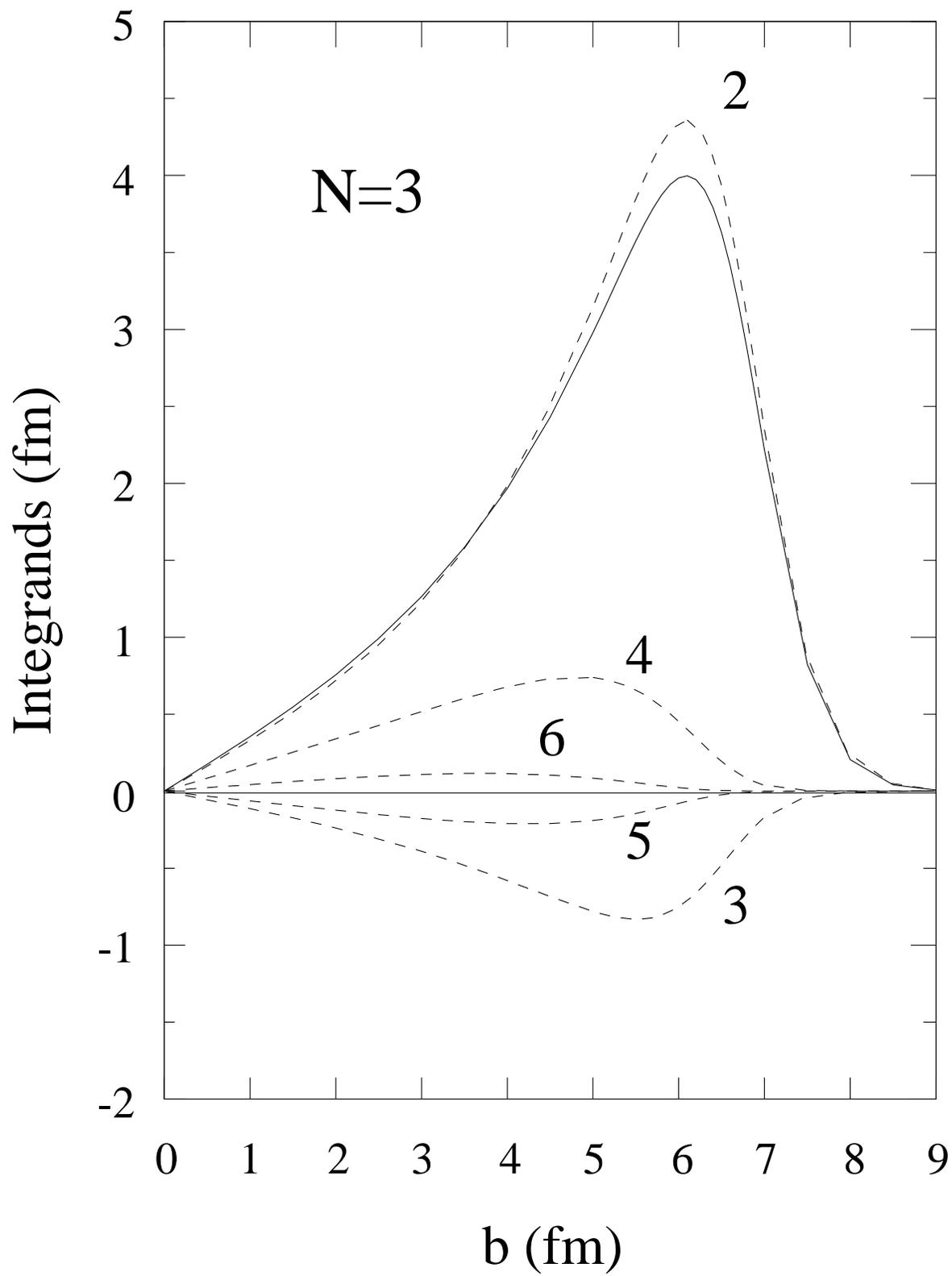
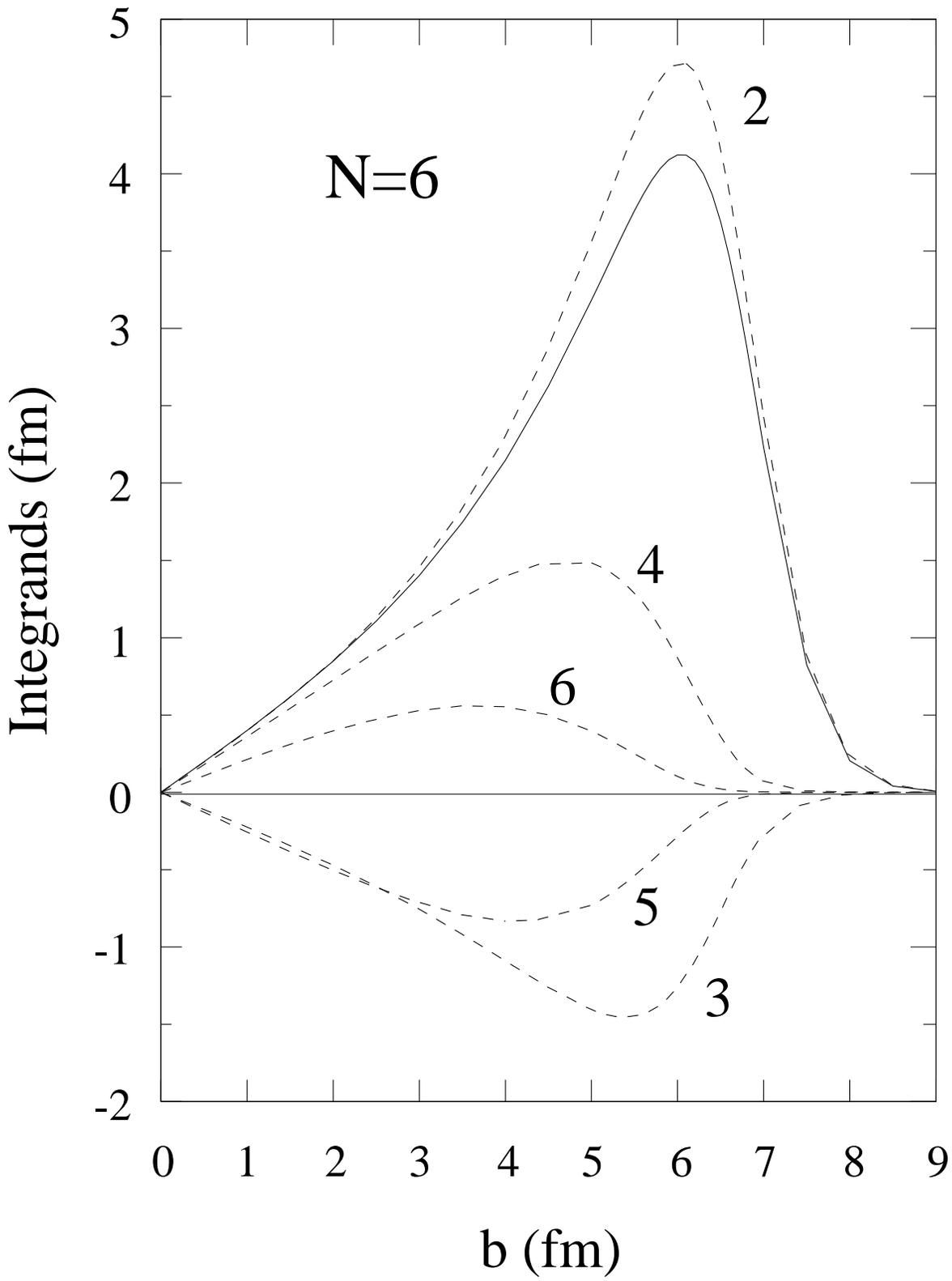
\end{document}